\documentclass[12pt]{article}

\usepackage{a4,epsfig,amsmath,amssymb,graphicx,multirow,mathrsfs}

\makeatletter
\renewcommand{\fnum@figure}{\footnotesize\textbf{\figurename~\thefigure}\,}
\renewcommand{\fnum@table}{\footnotesize\textbf{\tablename~\thetable}\,}
\makeatother

\newcommand{\sdot}{\!\cdot\!}
\newcommand{\SLASH}[1]{/\!\!\! #1}

\begin{document}
\title{ 
THE GROUND SCALAR NONET AND D DECAYS}
\author{
Olivier Leitner \\
{\em  Laboratori Nazionali di Frascati, Frascati} \\ 
{\em and Istituto Nazionale di Fisica Nucleare}\\ \\
Beno\^it Loiseau \\
{\em Laboratoire de Physique Nucl\'eaire et des Hautes \'Energies, Paris} \\ 
{\em groupe th\'eorie, IN2P3-CNRS, Universit\'es Paris VI-VII} \\ \\
 Jean-Pierre Dedonder    \\
{\em  IMNC/Universit\'e Paris VII, Denis-Diderot, Paris} \\ \\
Bruno El-Bennich \\
{\em Argonne National Laboratory,
High Energy Physics Division, Argonne} }
\maketitle
\baselineskip=11.6pt
\begin{abstract}
A short review on light scalar mesons is performed both in experiment and theory. 
A naive model, constrained by D branching ratios, is derived in order to make predictions 
on the wave functions of the $f_0(600)$ and $a_0(980)$ mesons. This leads us to compute 
transition form factors between the pseudoscalar $B$ and scalar mesons.   

\end{abstract}
\baselineskip=14pt
\section{What is a light scalar meson?}

Up to now, there is no global agreement on the interpretation of light 
mesons with vacuum quantum numbers: the scalar mesons\cite{penni}. At least, one can 
list two isovectors $a_0(980)$ and $a_0(1450)$, five isoscalars $f_0(600)/\sigma, 
f_0(980), f_0(1370)$, $f_0(1500)$ and $f_0(1710)$, and finally three 
isodoublets $K^{*}_0(800)/\kappa, K_0^{*}(1430)$ and $K^{*}(1950)$. One 
possible way to understand the light scalar spectrum may be to classify 
scalars according to their masses, i.e. below and beyond one GeV. Following 
this proposal, a first group with masses below one GeV (first nonet) contains 
$f_0(600)$, $K^{*}_0(800)$, $f_0(980)$ and $a_0(980)$. A second 
group with a mass beyond one GeV (second nonet) includes $f_0(1370), K_0^{*}(1430)$, 
$a_0(1450), f_0(1500)$, $f_0(1710)$ and $K^{*}(1950)$. Moreover, scalar 
mesons within their own group are built up according to the hypercharge, $Y$, 
and the isospin projection along the z-axis, $I_z$. The latter group being 
beyond the scope of this note, let us focus on the former group of light 
scalars so-called the first $SU(3)$ nonet.

\subsection{The first $SU(3)$ nonet}

\begin{figure}[H]
    \begin{center}
        {\includegraphics[scale=0.7]{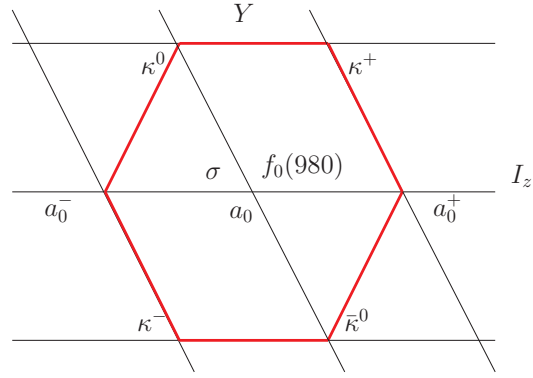}}
        \caption{\it The $SU(3)$ nonet}
\label{figsu3}
    \end{center}
\end{figure}

Following the spirit of the quark model, the $f_0(600)$ meson with quantum 
numbers $I^G(J^{PC})=0^+(0^{++})$, the $K^{*}_0(800)$ meson with quantum 
numbers $I^G(J^{P})=\frac12^+(0^{+})$, the $f_0(980)$ meson with quantum 
numbers $I^G(J^{PC})=0^+(0^{++})$ and the $a_0(980)$ meson with quantum 
numbers $I^G(J^{PC})=1^-(0^{++})$ constitute altogether the first scalar 
meson nonet given in fig.\ref{figsu3}. 
\noindent Regarding masses and widths, from the PDG\cite{pdg}, one has 
$M_{f_0(600)}=  400-1200 \,{\rm MeV}\ , $$\,$$ \Gamma_{f_0(600)}= 600-1000 \,{\rm MeV}$, 
$M_{K^{*}_0(800)}$$=  672\pm 40 \,{\rm MeV}\ , \, \Gamma_{K^{*}_0(800)}= 550\pm 34 \,{\rm MeV},$
$M_{f_0(980)}=  980\pm 10 \,{\rm MeV}\ , \Gamma_{f_0(980)}= 70\pm 30 \,{\rm MeV},$ and 
$M_{a_0(980)}=  985.1\pm 2.7 \,{\rm MeV}\ , \Gamma_{a_0(980)}= 75\pm 25 \;{\rm MeV}$, 
respectively. Various theoretical approaches in the study of different processes 
yield the following values for the pole of the $f_0(600)$\cite{theopol}:
\begin{align}
(489 \pm 26) - i(173 \pm 26)\ ,&  \;\;\; D^+ \to (\pi ^+\pi ^- )\pi ^+\ , \nonumber \\
(541 \pm 39) - i(252 \pm 42)\ ,&  \;\;\; J/\Psi \to \omega (\pi ^+\pi ^- )\ , \nonumber \\
(470 \pm 30) - i(295 \pm 20)\ ,&  \;\;\; \pi \pi \to \pi \pi \ , \nonumber
\end{align}
for the pole of the $K^{*}_0(800)$\cite{theopol}:
\begin{align}
(721 \pm 61) - i(292 \pm 131)\ ,& \;\;\; D^+ \to (K^-\pi ^+)\pi ^+\ ,\nonumber\\
(841 \pm 82) -i(309 \pm 87)\ ,& \;\;\; J/\Psi \to K^+\pi ^- K^-\pi ^+\ ,\nonumber\\
(722 \pm 60) -i(386 \pm 50)\ ,&\;\;\;K\pi \to K\pi\ , \nonumber
\end{align}
for the pole of the $f_0(980)$\cite{theopol}:
\begin{align}
(998 \pm 4) - i(17 \pm 4)\ ,&  \;\;\; J/\Psi \to \phi \pi ^+ \pi ^-\ , \nonumber \\
994 - i14\ ,& \;\;\; \pi \pi \to \pi \pi \;\;{\rm and}\;\; KK\ , \nonumber
\end{align}
and for the pole of the $a_0(980)$\cite{theopol}:
\begin{align}
(1036 \pm 5) - i(84 \pm 9)\ ,& \;\;\; \bar pp \to \eta \pi \pi \;\;
{\rm and}\;\; \omega \eta \pi ^0\ . \nonumber 
\end{align}
This non-exhaustive list of experimental and theoretical values underlines very
well the difficulties we have in understanding the structure and properties of
the scalar mesons.

\subsection{Experimental evidences of scalar mesons}

Unlike the difficulties to describe scalar mesons within a consistent 
theoretical framework, there are clear and unambiguous experimental 
evidences\cite{penni2} of light scalar mesons. Some of these indications also give 
crucial informations on their internal quark structure. 

\subsubsection{Observations}

Let us start with a few experimental signals provided by several 
collaborations. Regarding the  $f_0(600)$ meson\cite{sig}, which mainly 
decays into $\pi\pi$,  it has been observed in various processes. The 
phase shift of elastic $\pi\pi$ scattering,  when applying the Watson theorem 
and Roy equations, indicates the existence of $f_0(600)$. 
The E791 and FOCUS collaborations using isobar model (sum of Breit Wigner resonances) 
have also reported the $f_0(600)$ meson in $D^+ \to \pi^+ \pi^- \pi^+$ decay. 
Another way of observing $f_0(600)$ is related to the $p p \to p (\pi \pi) p$ central production
(GAMS collaboration) where a double pomeron ($\to \pi\pi$) governs the process at small momentum 
transfers between the protons. The BES and DM2 experiments have also noticed 
the $f_0(600)$ meson when the $\pi\pi$ angular distribution in 
$J/\psi \to f_0(600) \omega \to \pi \pi \omega$ was analyzed. For the 
$K^{\star}_0(800)$ meson\cite{kap}, which mainly decays into $K \pi$, 
two different analysis have drawn positive conclusions on its existence:  firstly,
the phase shift of elastic $K \pi$ scattering which was obtained from pion production 
by the LASS collaboration or from $D^+ \to K^- \pi^+ \pi^+$ by the FOCUS 
collaboration. Secondly, the E791 collaboration has also used an isobar model 
applied to $D^+ \to K^- \pi^+ \pi^+$ decay requires the $K^{\star}_0(800)$ for 
having a good fit of angular distributions. As regards the $f_0(980)$ 
meson\cite{f0}, which mainly decays into $\pi \pi$ 
and $K K$, two major observations have been made. The BES II collaboration in 
$J/\Psi \to \phi \pi^+ \pi^-$ and $J/\Psi \to \phi K^+ K^-$ decays has 
found prominent signals when data were fitted with a Flatt\'e formula. 
Another signal has also been observed in $D_s^+ \to \pi^- \pi^+ \pi^-$ decay by the 
E791 collaboration. The Dalitz plot analysis leads to suggest that a 
significant contribution is assumed to come from the $f_0(980)\pi^+$ channel 
and hence gives an experimental  evidence of the scalar $f_0(980)$.  Concerning 
the $a_0(980)$ meson\cite{a0}, which mainly decays into $\eta \pi$, 
one of the first signal was provided by the E852 collaboration using the $\pi^- p 
\to \eta \pi^+ \pi^- n$ reaction at $18.3 {\rm GeV}/{\rm c}^2$. The mass 
and width of the $a_0(980)$ meson were independently determined so that 
it gave a first clear signal of this scalar state.

\subsubsection{Quark structure}

The internal quark structure of light scalar is still 
controversial and only experimental observations can be used to test 
theoretical hypothesis\cite{penni2}. For example, let us consider here the case of 
$f_0(980)$ where several collaborations have confirmed the $s\bar{s}$ 
component of $f_0(980)$: the branching ratios (provided by the 
collaboration DM2 as well as by the PDG) of $\mathcal{B}r(J/\psi \to 
f_0(980) \phi)=(3.2 \pm 0.9) \times 10^{-4}$ and of $\mathcal{B}r(J/\psi 
\to f_0(980) \omega)= (1.4 \pm 0.5) \times 10^{-4}$ being different 
leads to a quark mixing in terms of $u\bar{u}$ and $s\bar{s}$ in  
$f_0(980)$. Finally, let us have a look at the $a_0(980)$ scalar for which
the collaboration KLOE\cite{kloe} has given the branching ratios for
radiative $\phi$ decays: $\mathcal{B}r(\phi \to \gamma f_0(980))=
(2.4 \pm 0.1) \times 10^{-4}$ and $\mathcal{B}r(\phi\to \gamma 
a_0(980))= (0.60 \pm 0.05) \times 10^{-4}$. The radiative decay 
$\phi\to \gamma a_0(980)$ which cannot proceed if $a_0(980)$ is a 
$\bar{q}q$ state can be however nicely described in the kaon loop
mechanism. This suggests a admixture of the $K\bar{K}$ component (4-quark state)
which is in contradiction with assuming $a_0(980)$ as a 2-quark state. 
Altogether, observing that   $a_0(980)$ and $f_0(980)$ are almost 
degenerate, one should have a $s\bar{s}$ component in $a_0(980)$ that 
cannot be since it is an $I=1$ state.

\subsection{Various theoretical models}

The fundamental structure of scalar mesons remaining unclear, together with
the difficulties related to experimental observation of the effects of light scalars in 
different processes, have generated a large variety of  theoretical models on the market, 
each of them claiming to explain the structure of light scalars below and 
beyond one GeV.  At least, five open-roads can be followed: the simplest 
one is the well-known $q\bar{q}$ state for describing light scalars, 
then the $q\bar{q}$ state plus glueball, then the four quark states 
$(qq)$$(\bar{q}\bar{q})$, and finally the mesonic 
molecules. Let us give a brief overview of their main characteristics\cite{mod}.
a) The $q\bar{q}$ state model where the $q\bar{q}$ L=0 nonet ($f_0(600), 
K^{*}_0(800)$, $a_0(980)$ and $f_0(980)$) is basically built up similarly to the $q\bar{q}$  
L=1 nonet ($\pi, \rho$...). This model however cannot explain why $a_0(980)$ 
and $f_0(980)$ are not degenerate, why the $a_0(980)$ and $f_0(600)$ have 
the same number of non strange quarks but are not degenerate, etc...
b) The  $q\bar{q}$ state plus glueball model where, according QCD expectations, 
the lightest glueball should be a scalar particle with quantum numbers 
$(J^{PC})=(0^{++})$. In such scenario, the glueball is considered as a 
very broad object with a width of the order of its mass.  It  works rather well 
for scalar particles with masses beyond one GeV. c) The four quark states 
$(qq)$$(\bar{q}\bar{q})$ model  which allows one to have two configurations 
in color space: $\bar{3}3$ and $6\bar{6}$. They can therefore rearrange to 
form a $(q\bar{q})(q\bar{q})$ scalar state.  d) Finally,
the mesonic molecule model which is similar to the $(qq)$$(\bar{q}\bar{q})$ 
case but considering only mesonic degree of freedom (color singlet) such 
as $\rho$ exchange for example.

\section{A toy model applied to the L=0 SU(3) nonet}

In our toy model, decay amplitudes for $D(D_s)$ to scalar and pseudoscalar 
mesons are evaluated by making use of the weak effective Hamiltonian at low 
energy together with QCD factorization. The associated  branching ratios are 
compared to the experimental ones. It leads to make predictions on transition form 
factors between pseudoscalar ($B$ and $D$) and scalar ($f_0(600)$ $K^*_0(800), 
f_0(980)$ and $a_0(980)$) mesons. We take advantage of these D decays to 
efficiently constrain, first the scalar meson wave functions and, then the transition 
form factors derived within a covariant relativistic formalism.

In Covariant Light Front Dynamics\cite{clfd}(CLFD), the state vector, 
which describes the physical bound state is defined on the light-front plane
given by the equation $\omega \sdot r = \sigma$. Here, $\omega$  denotes an 
unspecified light-like  four-vector ($ \omega^{2} = 0$) which  determines the 
position of the light-front plane and $r$ is a four-vector position of the 
system. Any four vector describing a phenomenon can be transformed from one 
system of reference to another one by using a unique standard matrix which 
depends only on kinematic parameters and on $\omega$. The particle is 
described by a wave function expressed in terms of Fock components of 
the state vector  which respects the properties required under any 
transformation. 

\subsection{Scalar wave functions}

 For a scalar particle composed of an  
antiquark and a quark of same constituent mass, $m$, the general 
structure of the two-body bound state has the form:                       
\begin{equation}\label{eq58.1}                                                    
\phi({\bf k}^2)=\frac{1}{\sqrt{2}}\bar{u}(k_2) A({\bf k}^2) v(k_1)\ ,
\end{equation}
where   $A({\bf k}^2)=N_S \exp \Bigl[- 4 \nu {\bf k}^2/m^2 \Bigr]$ 
is the  scalar component of the wave function. $N_S$ and $\nu$ are   
parameters to be determined from experimental $D$ branching ratios 
($D \to \, scalar \,\pi$ or $D \to \, scalar \, K$) and theoretical assumptions.  

\subsection{Transition form factors between pseudoscalar and scalar}

In CLFD, the approximate transition 
amplitude between a pseudoscalar, $P$, and a scalar, $S$, explicitly depends on the light front orientation:
\begin{equation}\label{eq9.7}
\langle S(P_{2}) | J^{\mu}| P(P_{1}) \rangle^{CLFD} = 
(P_{1}+P_{2})^{\mu} f_{+}(q^{2}) + (P_{1}-P_{2})^{\mu} f_{-}(q^{2})
     + B(q^{2})\omega^{\mu}\ , 
\end{equation}
where $B(q^{2})$ is a non-physical form factor which has to be zero in 
any exact calculation.  Simple algebraic calculations allow us to extract 
 the physical transition form factors 
$f{_{\pm}}(q^{2})$, by means of the amplitude 
$\langle S(P_{2}) | J^{\mu} | P(P_{1}) \rangle^{CLFD}$:
\begin{multline}\label{eq9.15}
\langle S(P_{2}) | J^{\mu} | P(P_{1}) \rangle^{CLFD}= \\
 \int_{(x,\tilde\theta,{\bf R}_{\perp})} D(x, \tilde\theta,{\bf R}_{\perp})
{\rm Tr} \biggl[ - {\Bar \vartheta_{S}} (m_1+ \SLASH k_{1})
 \gamma^{\mu}\gamma^5
(m_{2}+ \SLASH k_{2}) \vartheta_{P} (m_{3}- \SLASH k_{3}) \biggr]  \frac{1}{1-x^{\prime}}\ ,
\end{multline}
which is derived from the usual triangular diagram describing transitions between mesons.
$D(x, \tilde\theta,{\bf R}_{\perp})$ is the invariant phase space element and $\vartheta_{P}$ and $\vartheta_{S}$ denote respectively the initial 
pseudoscalar and final scalar wave functions. For more informations on the CLFD 
approach, we refer the reader to the paper\cite{clfd}.

\section{Conclusion}

\begin{figure}[H]
    \begin{center}
        {\includegraphics[scale=0.38]{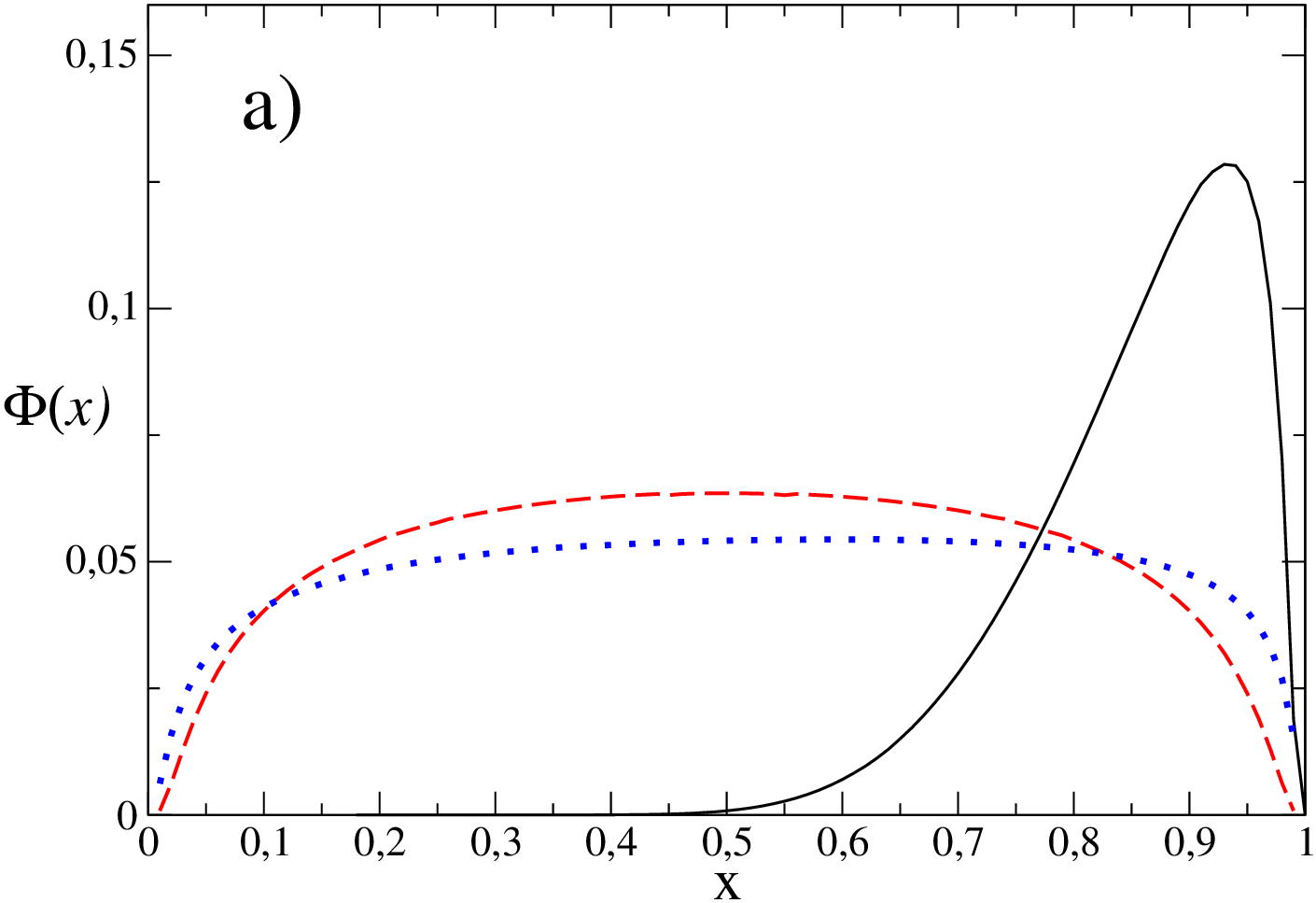}}
        \vskip 1.3cm
        {\includegraphics[scale=0.38]{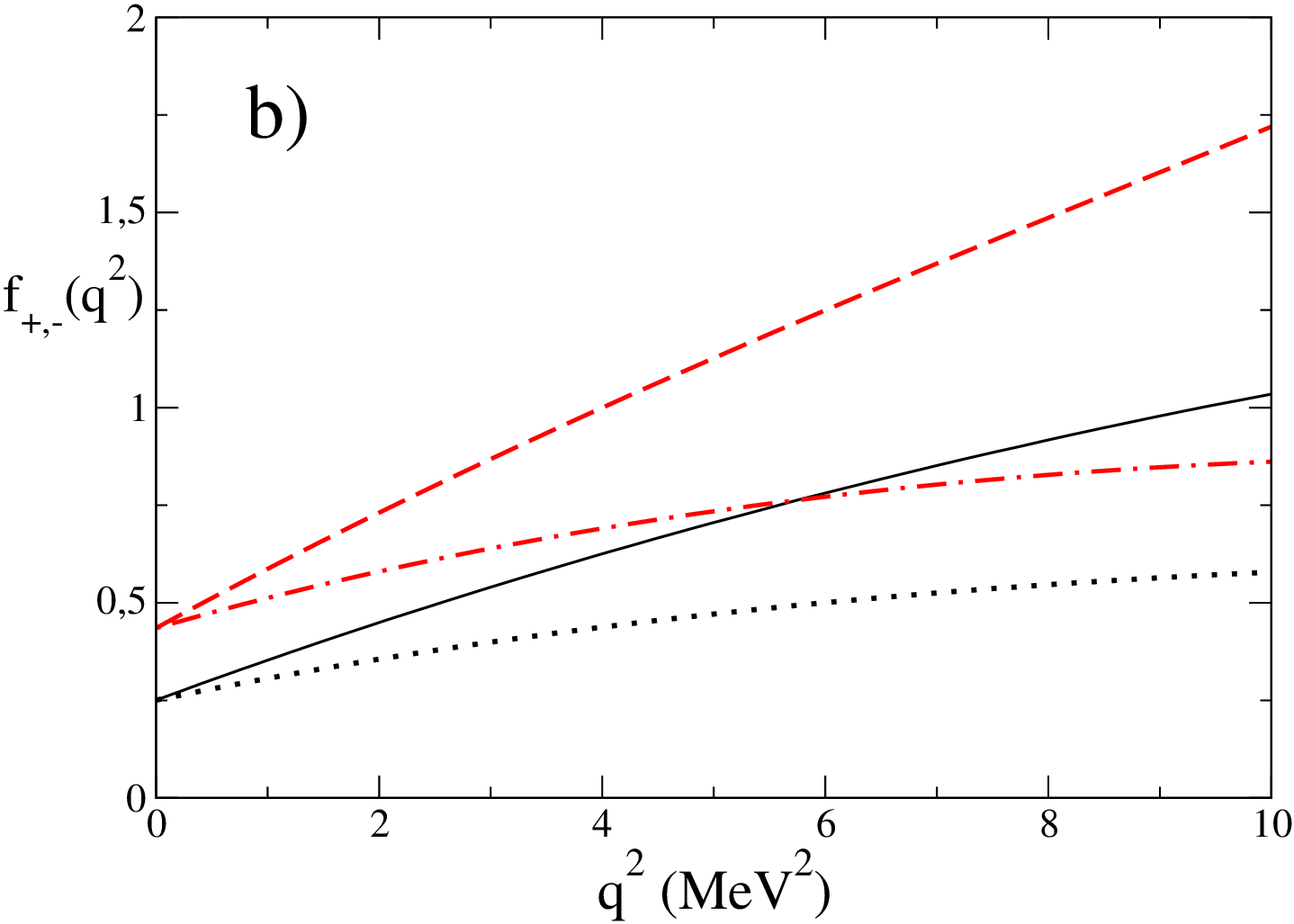}}
        \caption{\it a)  $x$-distributions for the $B$ meson (full line) 
as well as for $K^*(800)$ (dashed line) and $a_0(980)$  (dotted line). 
b) transition form factors, $f_{+}(q^2), f_{-}(q^2)$, plotted in 
case of $B \to f_0(600)$ (full and dotted lines) and $B \to a_0(980)$ 
(dashed and dotted-dash lines), respectively.}
\label{exfig}
    \end{center}
\end{figure}

Using normalization and $D$ experimental branching ratios one can model  
the wave function of scalar mesons for which some $x$ distributions are  
given in fig.~\ref{exfig}a). One can also make predictions on transition form 
factors between pseudoscalar and scalar mesons as shown (similar results for 
$D \to scalar$ transitions) in fig.~\ref{exfig}b). All the results given here 
are only qualitative due to some uncertainties among them the experimental 
$D$ branching ratios, the 2-quark description assumption and the meson and quark 
mass effects. It is therefore crucial to improve our understanding of scalar mesons 
as they play a major role when analyzing for example the $CP$ violation asymmetry in 
$B \to \pi \pi \pi(K)$. Better one knows the unitarity triangle, better one can 
 looks for new physics effects, however tiny they may be.

\end{document}